\documentstyle[aps,prl,amsmath,amssymb,multicol,graphicx]{revtex}
\renewcommand{\narrowtext}{\begin{multicols}{2}\global\columnwidth20.5pc}
\renewcommand{\widetext}{\end{multicols}\global\columnwidth42.5pc}
\title{Spin dynamics of the model 2D quantum antiferromagnet CFTD}
\author{H. M. R\o{}nnow$^{1,2}$, D. F. McMorrow$^1$, R. Coldea$^{3,4}$, A. Harrison$^5$, I. D. Youngson$^5$,\\
 T. G. Perring$^4$, G. Aeppli$^6$, O. Sylju\aa{}sen$^7$, K. Lefmann$^1$ and C. Rischel$^8$}
\address{
$^1$Condensed Matter Physics and Chemistry Department,Ris\o{} National Laboratory,
DK-4000 Roskilde, Denmark\\
$^2$DRFMC, CEA, 17, Rue des Martyrs, 38054 Grenoble, France\\
$^3$Solid State Division, Oak Ridge National Laboratory, TN 37831, USA\\
$^4$ISIS facility, Rutherford Appleton Laboratory, UK\\
$^5$Department of Chemistry, University of Edinburgh, The King's Buildings, Edinburgh EH9 3JJ, UK\\
$^6$NEC Research, 4 Independence Way, Princeton, NJ 08540, USA\\
$^7$NORDITA, Copenhagen, Denmark\\
$^8$Niels Bohr Institute, Copenhagen, Denmark}

\date{\today}

\begin{document}
\maketitle
\begin{abstract}
The magnetic excitation spectrum in the two-dimensional (2D) $S=1/2$ Heisenberg
antiferromagnet copper deuteroformate tetradeuterate (CFTD) has been measured for
temperatures up to $T\sim J/2$, where $J=6.31\pm0.02$~meV
is the 2D exchange coupling.
For $T\ll J$, a dispersion of the zone boundary energy is observed, which
is attributed to a wavevector dependent quantum renormalization.
At higher temperatures, spin-wave-like excitations
persist, but are found to broaden and soften.
By combining our data with numerical calculations, and with existing
theoretical work, a consistent description of the behaviour of the
model system is found over the whole temperature interval investigated.
\end{abstract}
\pacs{PACS numbers: }

\narrowtext

While at the atomic level, magnetism is a quantum phenomenon, the
collective behavior of magnets can be largely understood using
classical concepts. This can even be true for antiferromagnets with low
spin and spatial dimensionality, where quantum fluctuations are sizable.
A famous example is the 2D quantum ($S=1/2$) Heisenberg
antiferromagnet on a square lattice (2DQHAFSL) with nearest-neighbour
interactions.
Considerable effort has been devoted to
this particular model because it describes
the  parent compounds of the high-$T_c$
superconducting cuprates, and also because it once was thought to have
a spin fluid rather than a N\'eel ground state.

By now, it is well established
that at zero temperature the 2DQHAFSL displays long-range order,
albeit with a staggered moment reduced by quantum fluctuations
\cite{igarashi92}. In addition, harmonic spin-wave (SW) theory, based on a
classical image of the spins as coupled precessing tops, gives an
excellent account of the spin dynamics up to intermediate frequencies at
T=0. At finite temperatures,  long-range order is destroyed by thermal
fluctuations, and the system  possesses short-range order only,
characterized by a temperature-dependent correlation length $\xi(T)$.
A ombination of low temperature renormalised
classical theory \cite{chakravarty89}, intermediate temperature quantum
Monte Carlo (QMC) \cite{kim98,beard98} and high-temperature methods
\cite{elstner95b,cuccoli97} accounts for the experimental data for
$\xi(T)$, covering the range $J/5<T\lesssim
J$\cite{birgeneau99,greven95,ronnow99_cftd,carretta00}.
Thus, a coherent picture exists for the thermodynamic properties at
all temperatures for S=1/2, as well as for higher
values of the spin\cite{hasenfratz00}. The remaining questions about the
2DQHAFSL concern the intermediate and low frequency spin dynamics at
nonzero $T$, and the high frequency spin dynamics at all $T$. In both cases
one could well expect to see more severe quantum effects because
they should be more sensitive than the static properties to the
non-linearities of Heisenberg's equations for a low spin system. The
present paper describes the first experiment to confront all of these questions
directly over the full energy scale set by $J$.

Though experiments on the dynamics of the 2DQHAFSL are relatively
scarce, considerable theoretical work exists. Time dependent information is
contained in the dynamical structure factor
$S(Q,\omega)=
\int d\omega e^{-i\omega t}\sum_r e^{iQr}\langle S_0(0)S_r(t)\rangle$,
which is a function of wavevector $Q$ and energy $\hbar\omega$, and
is measured directly in neutron scattering experiments.
The $T=0$ properties can be described by
classical SW theory, but with quantities such as
the SW velocity
$v_s=Z_c\sqrt{2}Ja$, spin stiffness $\rho_s=Z_\rho J/4$, and
susceptibility $\chi_\perp=Z_\chi/(8J)$  renormalized by quantum
corrections.
Consistent values for the renormalization constants
$Z_c=1.18$, $Z_\rho=0.72$ and $Z_\chi=0.51$ have been obtained
using the
SW expansion to order
$1/(2S)^2$\cite{igarashi92}, and series expansion from
the Ising-limit\cite{singh95}.
However, these two approaches to the $T=0$ spin dynamics disagree
in detail, in that the SW expansion predicts
a constant energy at the zone
boundary (ZB), whereas the Ising-limit expansion predicts a $\sim$7\% dispersion.

On warming, scattering from thermally
excited magnons, and a variation of the order parameter
on a length scale set by $\xi$, will limit the lifetime of the excitations.
Correspondingly, the inverse magnon lifetime, $\Gamma$, will increase, while
the dispersion softens to lower energies.
Various theoretical studies
\cite{tyc89,auerbach88,takahashi89,sokol96,sherman99,winterfeldt97,nagao98,wang97} and QMC calculations in the limited temperature range
$0.35J<T<0.5J$\cite{makivic92} have addressed the dynamic structure
factor of the 2DQHAFSL at finite $T$. Most 
approaches agree on the existence of well-defined excitations, but few
experimental data are available to test the various quantitative predictions.
Much of the experimental work on the 2DQHAFSL has
focussed on La$_2$CuO$_4$\cite{birgeneau99} and
Sr$_2$CuCl$_2$O$_2$\cite{greven95,thurber97}, both of which contain the CuO$_2$
planes that are the building blocks of the cuprates. 
These materials have the drawback that their large coupling constant
$J\sim1500$~K makes studies on temperature and energy scales
comparable to $J$ technically challenging. 
Another realization of the 2DQHAFSL
is Cu(DCOO)$_2\cdot$4D$_2$O (CFTD), where the $S=1/2$ Cu moments are
coupled through formate groups, leading to
an exchange energy ($J=73.2$~K) which is more amenable to experiments.
The correlation length $\xi(T)$ in
CFTD has recently been measured up to
temperatures of $T\sim J$ \cite{ronnow99_cftd,carretta00},
and found to be in good agreement with theory and computations.

Below 236 K, CFTD has the $P2_1/n$ space group with lattice parameters
$a=8.113$~\AA, $b=8.119$~\AA, $c=12.45$~\AA{}, and a monoclinic angle
$\beta=101.28^\circ$.
The $ab$ plane contains face centred $S=1/2$ Cu$^{2+}$ ions,
forming an almost square lattice.
The SW dispersion along $a^\ast$ and $b^\ast$
measured by neutron scattering is well described by an isotropic
nearest-neighbour coupling $J=6.31$~meV, and a small
anisotropy induced gap of $0.38$~meV at the zone centre\cite{clarke99}.
The inter-layer coupling is estimated to be less than $10^{-4}J$,
while a  Dzyaloshinskii-Moriya term $J_D=0.46$~meV has been
inferred from magnetization measurements\cite{yamagata81}.
Below $T_N=16.54$~K the system orders three-dimensionally due to the
inter-layer coupling \cite{burger80}.

Our neutron inelastic scattering experiments using the direct geometry
time-of-flight (TOF) spectrometer HET at ISIS, UK.
The incident energy $E_i=25$~meV gave an energy resolution of FWHM 1.64~meV
at zero energy transfer.
A 3.71~g single crystal of CFTD was aligned with the 2D planes normal
to the incident beam. In the geometry used the horizontal and vertical detector
banks measured scattering along the [1,1]-type direction, while the
diagonal banks collected data along the [1,0] directions.
The measured intensities were corrected for detector efficiency, and
converted to absolute units by normalising to the incoherent
scattering from a vanadium standard.

\begin{figure}
\noindent\hspace{-1mm}\includegraphics[width=0.48\textwidth]{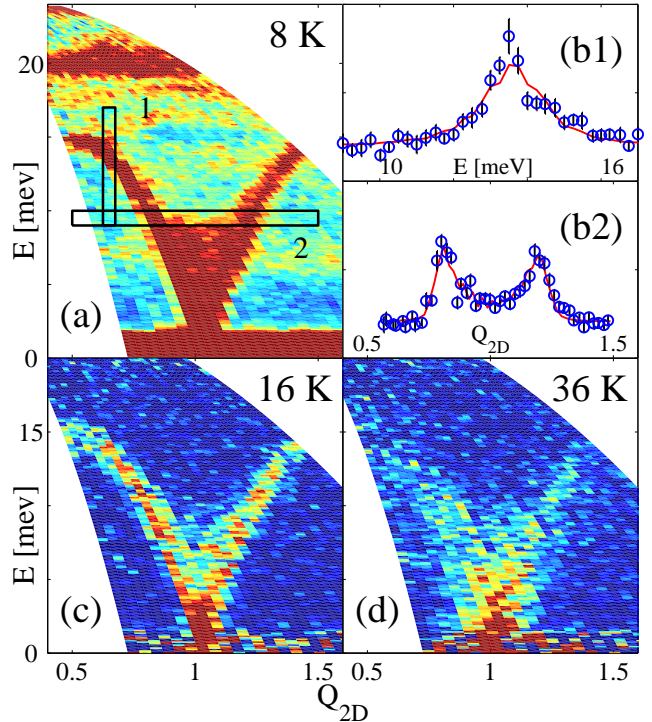}
\caption{Inelastic TOF data in CFTD with $Q_{2D}$ in units of $(\pi,\pi)$.
(a) Raw data at 8~K. (b) Representative cuts along energy (b1) and $Q$ (b2) as
indicated by the windows in panel (a).
The vertical scales cover 0 to 100~mbarn/steradian/meV/spin, and
the solid lines are fits to the resolution convoluted cross-section
described in the text.
(c)--(d)
Background subtracted data at
16~K and 36.2~K  respectively.
The pseudo-colour scale (from blue to dark red)
range from 0 to 50~mbarn/steradian/meV/spin.}
\label{fig:rawdata}
\end{figure}

Data were collected at several temperatures between 8~K
and 150~K with typically 24 hours of counting per temperature.
The data at 8~K (within the 3D ordered phase)
provide a precise determination of the spin Hamiltonian, and
allow characterization of the phonon background.
For example, the
band at 20~meV is probably due to the
localised motion of the crystal bound
water, which freezes in an anti-ferroelectric transition at
236~K~$\sim$~20.3~meV\cite{burger80}, 
while the extra scattering around 7~meV is
due to acoustic phonons emerging from the (1,0,1) type
reciprocal lattice points, see Fig. \ref{fig:rawdata}(a).

At 8~K, the magnetic signal is concentrated along a sharp SW
dispersion curve. The data were analysed by taking cuts
along $Q$ and  $\omega$, as illustrated in Fig.\ \ref{fig:rawdata}(b),
allowing the phonon contribution to be dealt with as a local background.
Each cut was fitted with a parameterized model for the scattering obtained
by convolving the full instrumental resolution with
the linear SW theory form $S(Q,\omega)=
\frac{A}2\sqrt{\frac{1-\gamma_Q}{1+\gamma_Q}}\delta(\omega-\omega_Q)$;
$\omega_Q=2\tilde{J}\sqrt{1-\gamma_Q^2}$;
$\gamma_Q=\frac12(\cos Q_x+\cos Q_y)$,
multiplied by the magnetic form factor of free Cu$^{2+}$ and the Bose
population factor.
Since the parameterization is only applied locally 
(the parameters $A$ and $\tilde{J}$ were allowed to vary for each cut),
it imposes no constraints on the global form of $S(Q,\omega)$ or on
the extracted dispersion, shown in Fig.\ 2.
It is evident that along [1,1] the
dispersion is well described by a uniform renormalization
of the linear SW theory result. A fit gave
$J=\tilde{J}/Z_c=6.31\pm0.02$~meV, in good agreement with previous
neutron scattering \cite{clarke99} studies and the value derived from
high-temperature susceptibility \cite{yamagata81}.
The present data can be combined with the susceptibility data
to give a value of the
renormalization factor $Z_c=1.21\pm0.05$, within error the
same as the theoretical value 1.18 \cite{igarashi92,singh95}.
\begin{figure}
\includegraphics[width=0.45\textwidth]{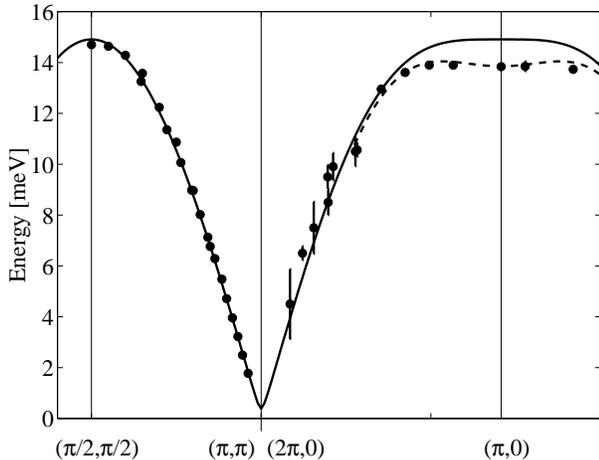}
\caption{The SW dispersion in CFTD at 8~K.
The solid line is a fit of the $Q||[1,1]$ part to nearest neighbour
SW theory, giving $J=6.31\pm0.02$~meV.
The dashed line is the result of an expansion from the Ising-limit
\protect\cite{elstner95b} using $J=6.31$~meV.}
\label{fig:disp}
\end{figure}

Along [1,0] there is a pronounced dip in energy
around $(\pi,0)$, indicating a dispersion of $6\pm1$\% along
the zone boundary (ZB) from $(\pi/2,\pi/2)$ to $(\pi,0)$.
Such a ZB dispersion has also been observed
in Sr$_2$Cu$_3$O$_4$Cl$_2$ \cite{kim99},
where it was explained in terms of the Ising-limit expansion
\cite{elstner95b}.
Indeed, the Ising-limit expansion accounts perfectly for our
data without any additional parameters (dashed line in Fig.\
\ref{fig:disp}).
However, our  data can equally well be modelled within SW
theory by introducing an antiferromagnetic next-nearest-neighbour
coupling of $(0.067\pm0.007)\times J$, and
{\it a priori} the experimental data cannot be used to
discriminate between
these two possibilities.
To help resolve this issue we have undertaken
numerical computations on finite sized systems, 
and do indeed find a ZB dispersion for
pure nearest neighbour coupling \cite{ronnow_unpublished}.
QMC on lattices up to $32\times32$ spins and temperatures down to
$0.3J$ extrapolate to about 6\% ZB dispersion \cite{syljuasen00}.
Exact diagonalisation of an $8\times8$ system display a ZB dispersion
of 4.8\%. Both techniques yield no ZB dispersion in the $4\times4$
system, which indicate that a significant number of spins are involved
in this effect. We conclude that the experimentally observed ZB dispersion is
indeed an intrinsic quantum effect on the $T=0$ spin dynamics of the
pure 2DQHAFSL model with nearest-neighbour interactions.
Recently a ZB dispersion of $-13$\%  (i.e. opposite sign)
has been found in La$_2$CuO$_4$, which is
attributed to higher-order spin couplings \cite{coldea00b}.

Above $T_N$, the magnetic excitation spectrum broadens, making it
more difficult to distinguish from the phonon scattering. To determine
the phonon scattering, the SW contribution was simulated by
convolving the linear SW result with the experimental resolution. When
subtracted from the 8~K data the remaining intensity is dominated by
phonon scattering. This was scaled to the thermal
population factor and subtracted from the remaining $T>T_N$ data.
As this method assumes a uniformly renormalised SW dispersion, it
obviously does not apply to $Q||[1,0]$, and hence only the $Q||[1,1]$
data were analysed in this way.
Fig.\ 1(c)-(d) shows examples of data sets where the phonon
contributions have been removed.
The background corrected data were analysed by fitting cuts along
energy to the same parameterization as for the 8 K data, but with the delta
function replaced by the damped harmonic oscillator (DHO) line shape
$\delta(\omega-\omega_Q)\rightarrow\frac4\pi\frac{\Gamma\omega_Q\omega}
{(\omega^2-\omega_Q^2)^2+4\Gamma^2\omega^2}$.
No clear trends could be found within error for the $Q$ dependence of the
fitting parameters $A$, $\tilde{J}$ and $\Gamma$. In the following we
therefore choose to discuss the $Q$ averaged
values $\bar{A}(T)$, $\bar{J}(T)$ and $\bar{\Gamma}(T)$.

\begin{table}
\begin{tabular}{l|rrrrrrr}
$T$~[K] & 8.0 & 16.2 & 20.9 & 25.6 & 30.4 & 36.2 & 45.1\\
$Z_\chi$ & 0.51(4) & 0.49(2) & 0.51(2) & 0.49(2) & 0.42(3) &
0.58(5) & 0.41(8)\\
$Z_c$ & 1.18& 1.19(1)&1.17(1)&1.17(1)&1.14(2)&1.07(4)&1.03(4)\\
$\bar{\Gamma}/J$&0.00(2)& 0.08(1) & 0.12(1) & 0.17(2) & 0.24(3) &
0.38(5) & 0.42(7)
\end{tabular}
\caption{Temperature dependence of the susceptibility and SW
velocity renormalization parameters, $Z_\chi$ and $Z_c$, and the
SW damping $\bar{\Gamma}(T)/J$.
$Z_c$ has been normalised to match the theoretical value 1.18 at 8~K.
The 8~K data were resolution limited, and therefore correspond to zero
SW damping.}
\label{tab:tdep}
\end{table}

The amplitude, $\bar{A}(T)=\frac23(\frac{\gamma e^2}{mc^2})^2Z_\chi S$,
can be used to extract the renormalization $Z_\chi(T)$ of the susceptibility.
(The factor $\frac23$ becomes $\frac12$ for the horizontal banks
below $T_N$ where the moments order almost vertically.)
The temperature dependence of  $Z_\chi(T)$ is listed in Table
1. At low $T$, the experimental value is in perfect agreement with
the predicted value of 0.51 \cite{igarashi92}.
At higher $T$, $Z_\chi$ appears to decrease, but no clear predictions
for the $T$ dependence of this quantity have yet been reported.

The uniform SW renormalization of $\omega_Q$ found along [1,1] at 8~K
justifies the use of the average $\bar{J}(T)$, which divided by $J$
gives the temperature dependent SW renormalization
$Z_c(T)$ as shown in Fig.\ 3.
Kaganov and Chubukov \cite{elstner95b,kaganov88}
have calculated the temperature dependence of the
higher-order quantum corrections to SW theory, obtaining
$v_s(T)=v_s(0)\left(1+\frac{\zeta(3)}{4\pi}\left(\frac{T}{JS}\right)^3\right)^{-1}$,
where $\zeta(3)\simeq1.2021$.
This result is represented by the solid line and
is in good agreement with our data.
The new QMC results are indicated by triangles  and
are also consistent with the data.

\begin{figure}
\center
\includegraphics[width=0.45\textwidth]{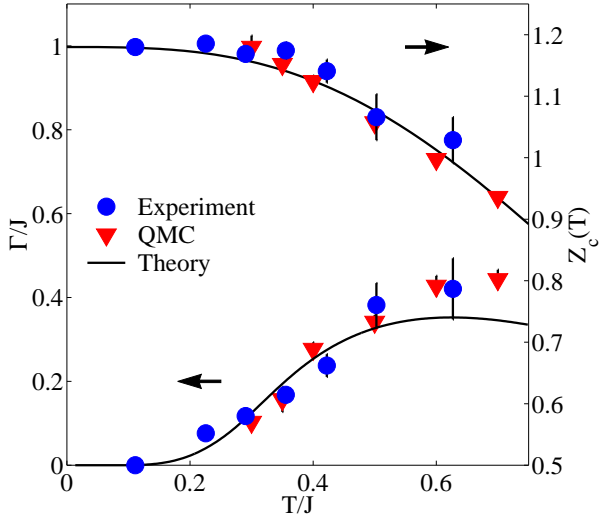}
\caption{The SW softening $Z_c=\tilde{J}/J$ and damping $\Gamma/J$
as functions of temperature.
The experimental data (circles) and new QMC results (triangles) are
averages for $\omega>2$~meV.
The solid
lines are respectively the prediction of Kaganov and Chubukov
\protect\cite{elstner95a,kaganov88} for $Z_c(T)$ and the relation
$\Gamma(T)=v_s(T)/\xi(T)$ for the damping.}
\label{fig:gamma_soft}
\end{figure}

Various calculations of the SW damping
all predict a $Q$ dependence with a minimum at
$(\pi,\pi)$. Around this point, the value of  $\Gamma$ is quite
dependent on the definition of the line shape. This complication is
irrelevant for the experimental data, where due to the large incoherent
background, the fits were restricted to $\omega>2$~meV~$\simeq J/3$.
The experimentally determined SW damping is shown in
Fig. \ref{fig:gamma_soft}.
QMC results for the damping
have been reported for $0.35<T/J<0.5$ \cite{makivic92}.
We find that these overestimate
the experimental values of $\bar{\Gamma}$
by almost a factor of two.
To resolve this discrepancy, we have repeated the QMC
calculations \cite{ronnow_unpublished}. If the same
maximum entropy (ME) method is used
for analytic continuation of the
imaginary time MC data, then the results of Ref. \cite{makivic92} are
reproduced.
We found, however, that it is more robust to
impose the same DHO line shape as used to analyse our
experimental data. When this is done the
QMC values  shown in Fig.\
\ref{fig:gamma_soft} are obtained which are in
good agreement with the experimental data.
Scaling arguments have also been used to
calculate the SW damping \cite{tyc89}.
Since this approach mainly applies to the low-energy, long-wavelength
behaviour of the system which was not probed in our experiments,
we believe that it is
not appropriate to compare it with our data.
We do note, however, that for
$T< J/2$ the data lie above the prediction for
$Q=(\pi,\pi)$, and below the prediction extrapolated to
$Q=(\pi/2,\pi/2)$.
Our result for $\Gamma(T)$ differ from those extracted from 
NMR in Sr$_2$CuCl$_2$O$_2$ \cite{thurber97}. This may be becase NMR is a
local probe in the low energy limit, and the result for
$\Gamma(T)$ relies on a global asumption for $S(Q,\omega)$. In
contrast, our data give a direct and asumption free measurement of
$\Gamma(T)$.

Spin waves are eigenmodes with respect to antiferromagnetic
order and vary on a  length scale set by the correlation length
$\xi(T)$.
The most naive way to estimate the
lifetime of a SW would be to divide $\xi(T)$ \cite{hasenfratz00}
by the SW velocity $v_s(T)$ \cite{elstner95a,kaganov88}.
In Fig.\ 4 the solid
line is the inverse lifetime $\Gamma(T)=v_s(T)/\xi(T)$
obtained in this way, and is seen to be in surprisingly good
agreement with the data.

In summary, we have measured the excitation
spectrum of CFTD, which is an excellent physical realization
of the 2DQHAFSL.
Combining our data and numerical calculations,
the existence of a quantum
induced ZB dispersion has been unambiguously established.
The finite temperature behaviour has been probed up
to $T>J/2$, where well-defined SW like excitations persist.
The temperature dependence of the SW softening and damping in this
strongly quantum mechanical system are remarkably well described,
without any adjustable parameters, by existing
theories based on a quantum renormalization of the classical system.

We gratefully acknowledge Rajiv R. P. Singh for
useful discussions. This work was supported by the Danish Research Academy, the
UK EPSRC and the EU through its TMR and IHP programmes.
ORNL is managed for the US DOE by UT Battle, LLC, contract DE-AC05-00OR22725.

\bibliographystyle{prsty}
%\bibliography{../../../../phd/cftd/cftd,../../../../phd/mypublications}

\widetext
\end{document}